# Etude cinétique de CVD de pyrocarbone obtenu par pyrolyse de propane

ZIEGLER-DEVIN Isabelle, FOURNET René, MARQUAIRE Paul-Marie*

Département de Chimie Physique des Réactions (DCPR, UMR 7630 CNRS), Nancy Université
ENSIC, 1 rue Grandville, 54000 Nancy, France

## Résumé

La pyrolyse du propane à haute température (900-1000°C) et basse pression (<< 1 atm) produit un dépôt de pyrocarbone, mais surtout de l'hydrogène et des hydrocarbures allant du méthane à des espèces polyaromatiques (pyrène…). Une étude expérimentale a permis de quantifier 30 produits de la réaction dans différentes conditions opératoires. Un modèle détaillé de pyrolyse (600 réactions) a été élaboré et validé sur l'ensemble de ces résultats, il comprend un mécanisme homogène, décrivant les réactions en phase gazeuse de pyrolyse du propane, couplé à un mécanisme hétérogène, décrivant les réactions de dépôt du pyrocarbone.

**Mots-clés :** mécanisme, pyrolyse, propane, dépôt de pyrocarbone, CVD, CVI

## 1. Introduction

Les matériaux composites Carbone/Carbone sont utilisés dans les domaines aéronautique et spatial en raison de leurs propriétés : faible densité (4 fois inférieure à celle de l'acier), capacité calorifique importante, caractéristiques de friction, tenue à haute température, résistance à l'ablation, ce qui permet des usages en conditions extrêmes (freins pour avion, tuyères de moteur fusée, …). Les composites C/C sont produits par CVI ("Chemical Vapor Infiltration") de pyrocarbone dans un substrat poreux (préforme) constitué de fibres de carbone. Le dépôt de pyrocarbone est obtenu par pyrolyse d'hydrocarbures à haute température (vers 1000°C) et basse pression (<< 1 atm). La durée de l'infiltration est de plusieurs centaines d'heures et cette opération constitue un élément important du coût de fabrication des composites C/C. La compréhension et la modélisation des phénomènes chimiques et physiques qui interviennent lors du dépôt de pyrocarbone sont essentielles pour permettre d'améliorer le procédé, de maîtriser la qualité des produits et de réduire les coûts de production.

La pyrolyse d'hydrocarbures à haute température (le propane dans cette étude) produit le pyrocarbone recherché en faible proportion (<10%), mais également et majoritairement de l'hydrogène et de nombreux hydrocarbures allant du méthane à des espèces aromatiques et polyaromatiques. La formation de pyrocarbone a fait l'objet de nombreuses études (Bokros, 1965 ; Tesner, 1984 ; Becker et Hüttinger, 1998 ; Feron et al., 1999 ; Descamps et al., 2001 ; Vignoles et al., 2004…) ; des modèles globaux ou détaillés ont été proposés mais sont sujet à discussion et en particulier, la nature des précurseurs du pyrocarbone est toujours l'objet de controverses.

L'objectif de cette étude est de développer un modèle détaillé de la réaction comportant un mécanisme homogène basé sur des processus élémentaires, décrivant les réactions en phase gazeuse de pyrolyse du propane, couplé à un mécanisme hétérogène, décrivant les réactions de dépôt du pyrocarbone. Une étude expérimentale détaillée, en dosant le maximum de produits de la réaction dans un large domaine expérimental, est réalisée en réacteur parfaitement agité afin de contrôler au mieux les conditions opératoires. Pour s'affranchir des limitations diffusionnelles, cette étude est réalisée en conditions de CVD (Chemical Vapor Deposition).

* Auteur à qui la correspondance devrait être adressée : paul-marie.marquaire@ensic.inpl-nancy.fr

## 2. Etude expérimentale

Le pyrocarbone est produit par pyrolyse de propane dilué dans de l'azote à basse pression (≈2,7 kPa), et haute température (900-1000°C). Cette pyrolyse est réalisée dans un réacteur auto-agité par jets gazeux (RPA) (Matras et Villermaux, 1973) qui fonctionne en régime permanent, isotherme, isobare et isochore ; le temps de passage est de l'ordre de la seconde (0,5-5s). Il est réalisé en quartz et se compose de deux parties (dites « haute » et « basse ») qui se connectent l'une dans l'autre (Figure 1). La partie dite « haute » comporte une partie annulaire, qui constitue la zone de préchauffage et permet d'amener les réactifs à une température proche de celle de la pyrolyse en un temps court comparé au temps de réaction (Barbé et al., 1996 ; Ziegler et al., 2005a). Cette zone de préchauffage annulaire conduit directement à la croix d'injection, qui est composée de quatre tuyères qui produisent un jet dans des directions différentes et assurent le brassage de tout le volume du réacteur. Le réacteur contient dans sa partie dite « basse », un support circulaire amovible permettant de positionner les substrats en fibres de carbone utilisés pour réaliser les dépôts. Le support circulaire est percé en son centre, permettant ainsi de réaliser un prélèvement des effluents gazeux à proximité des substrats. Le réacteur mesure environ 90 cm de long, pour une zone réactionnelle de volume 87 $cm^3$ et de surface 100 $cm^2$.
Contrairement aux autres réacteurs (tubulaires) utilisés dans les différentes études (Becker et Hüttinger, 1998 ; Vignoles et al., 2004) sur la formation du pyrocarbone, le RPA utilisé dans cette étude présente l'avantage d'avoir la même composition gazeuse dans tout son volume ; il permet ainsi de connaître la concentration des différents produits issus de la pyrolyse du propane à proximité de la surface où se font les dépôts. La composition de cette phase gazeuse est un élément clé pour la compréhension des mécanismes de formation du pyrocarbone.

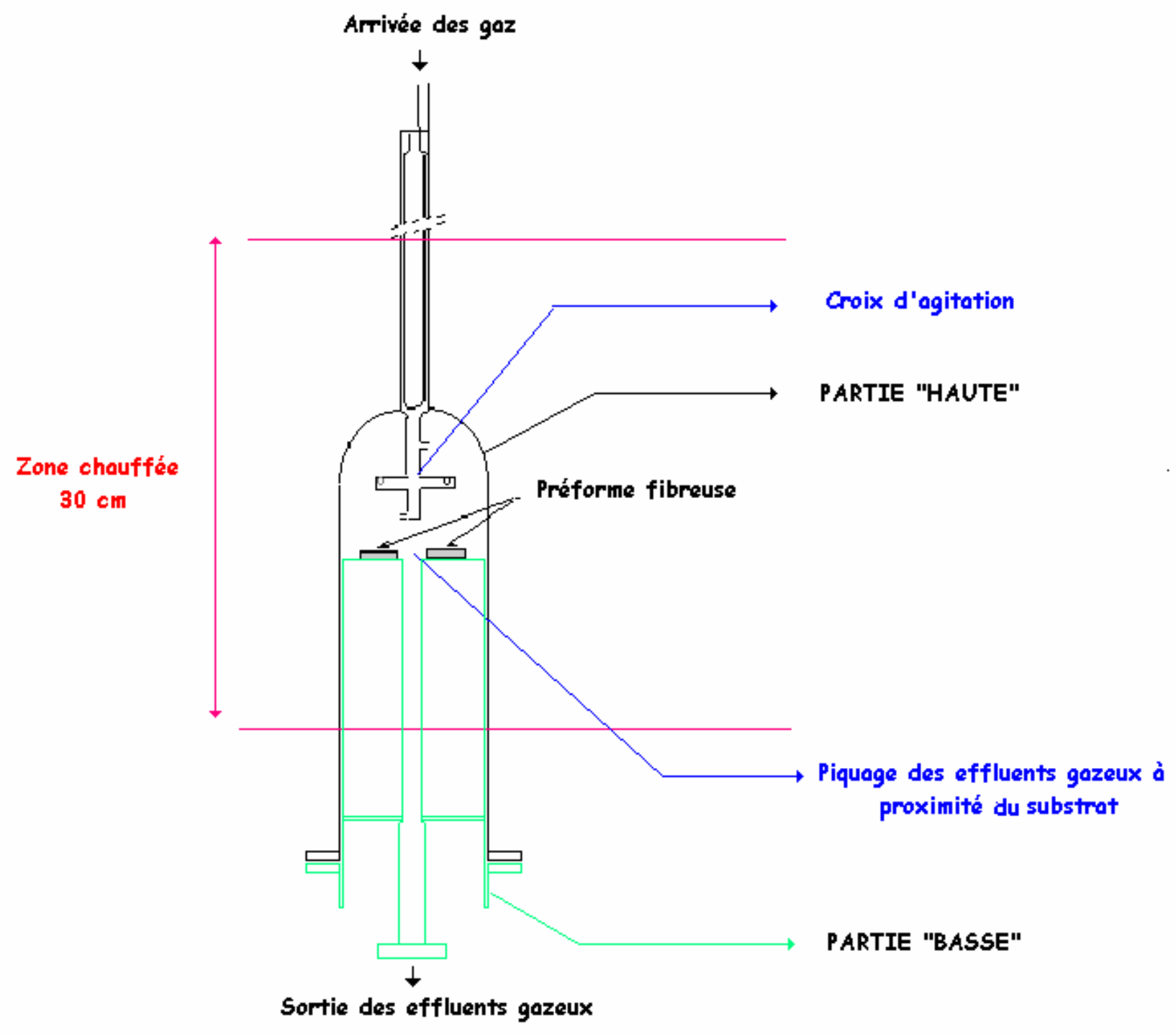

*Figure 1. Schéma du réacteur auto-agité*

Le montage expérimental permet de faire varier différents paramètres de la pyrolyse : la température, le temps de passage, la surface de la préforme,…

Concernant, les produits issus de la pyrolyse du propane, on distingue deux « catégories » de produits : les produits gazeux et les produits solides/liquides dans les conditions standards de température et de pression. Deux protocoles analytiques sont donc mis en place pour doser ces espèces. Tout d'abord, parmi les produits de pyrolyse, les produits légers ($H_2$, $C_1$-$C_4$) sont prélevés en sortie de réacteur et sont quantifiés (12 espèces) en ligne, par chromatographie en phase gazeuse ; l'échantillon gazeux est ainsi analysé à la même pression que celle régnant dans le réacteur. La répartition des produits légers, qui sont majoritaires, est donnée dans la Table 1. Dans un second temps, les produits plus lourds ($\geq C_5$) sont condensés dans un piège connecté à la sortie du réacteur et refroidi à l'azote liquide. L'analyse de ces condensats par spectrométrie de masse a montré la présence de plus de 50 espèces aromatiques et polyaromatiques. Parmi ces nombreuses espèces, 17 molécules en quantité importante, allant du cyclopentadiène au pyrène ($C_{16}$), sont quantifiées par chromatographie en phase gazeuse sur colonne capillaire. La quantification du dépôt est effectuée par pesée de chaque préforme fibreuse avant d'être introduite dans le réacteur et après la pyrolyse ; la différence de masse peut être assimilée au dépôt de pyrocarbone dans la préforme. Au total, on quantifie ainsi 30 produits de la réaction ; le détail de ces résultats a déjà été publié (Ziegler et al., 2005b ; Ziegler et al., 2005c).

*Table 1. Composition de la phase gazeuse (produits légers) Pyrolyse du propane dilué dans l'azote (2,7 kPa) à 1000°C et $\tau = 1$ s*

| produits | fraction molaire |
|---|---|
| $H_2$ | 1.31E-01 |
| $CH_4$ | 5,57E-02 |
| $C_2H_4$ | 4.43E-02 |
| $C_2H_2$ | 3.45E-02 |
| $pC_3H_4$ | 7.33E-04 |
| $aC_3H_4$ | 2.54E-04 |
| $C_3H_6$ | 3.03E-04 |
| $1,3\text{-}C_4H_6$ | 2.64E-04 |
| $C_4H_4$ | 6.90E-04 |
| $C_4H_2$ | 3.66E-04 |

Dans nos conditions, la conversion du propane est toujours totale et la prise de masse est proportionnelle avec la durée du dépôt de pyrocarbone (sur des durées de dépôt inférieures à 50 h) (Figure 2) ce qui indique qu'il n'y a pas de limitation diffusionnelle (faible infiltration) et donc que la réaction est contrôlée par la cinétique chimique. La vitesse de dépôt de pyrocarbone augmente avec la température (Figure 3) mais reste lente, environ 36 nm/h à 1000°C, et la fraction de pyrocarbone ne représente qu'une faible proportion (< 10%) des produits issus de la pyrolyse du propane.

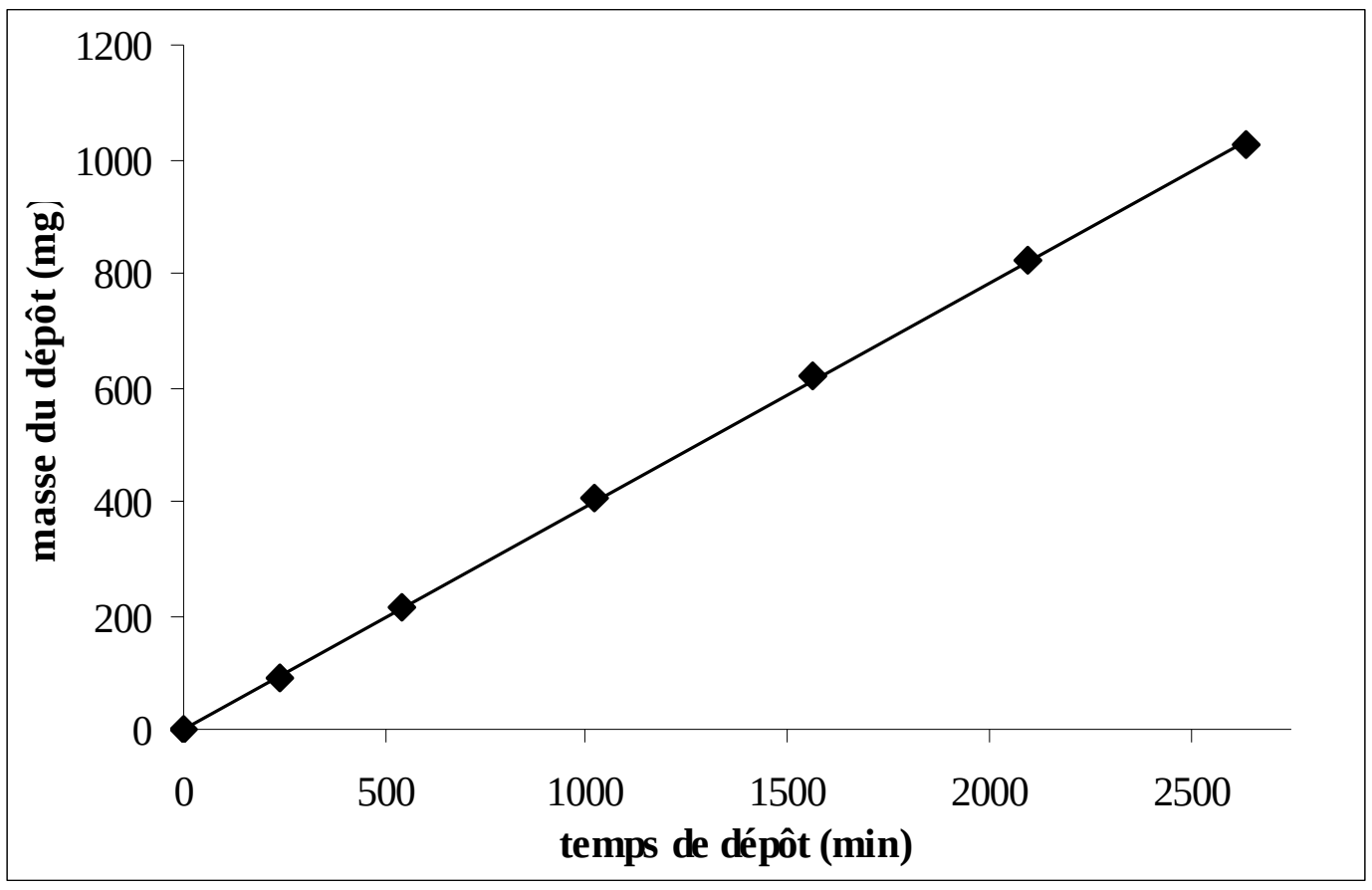

*Figure 2. Evolution de la prise de masse en fonction de la durée de dépôt*

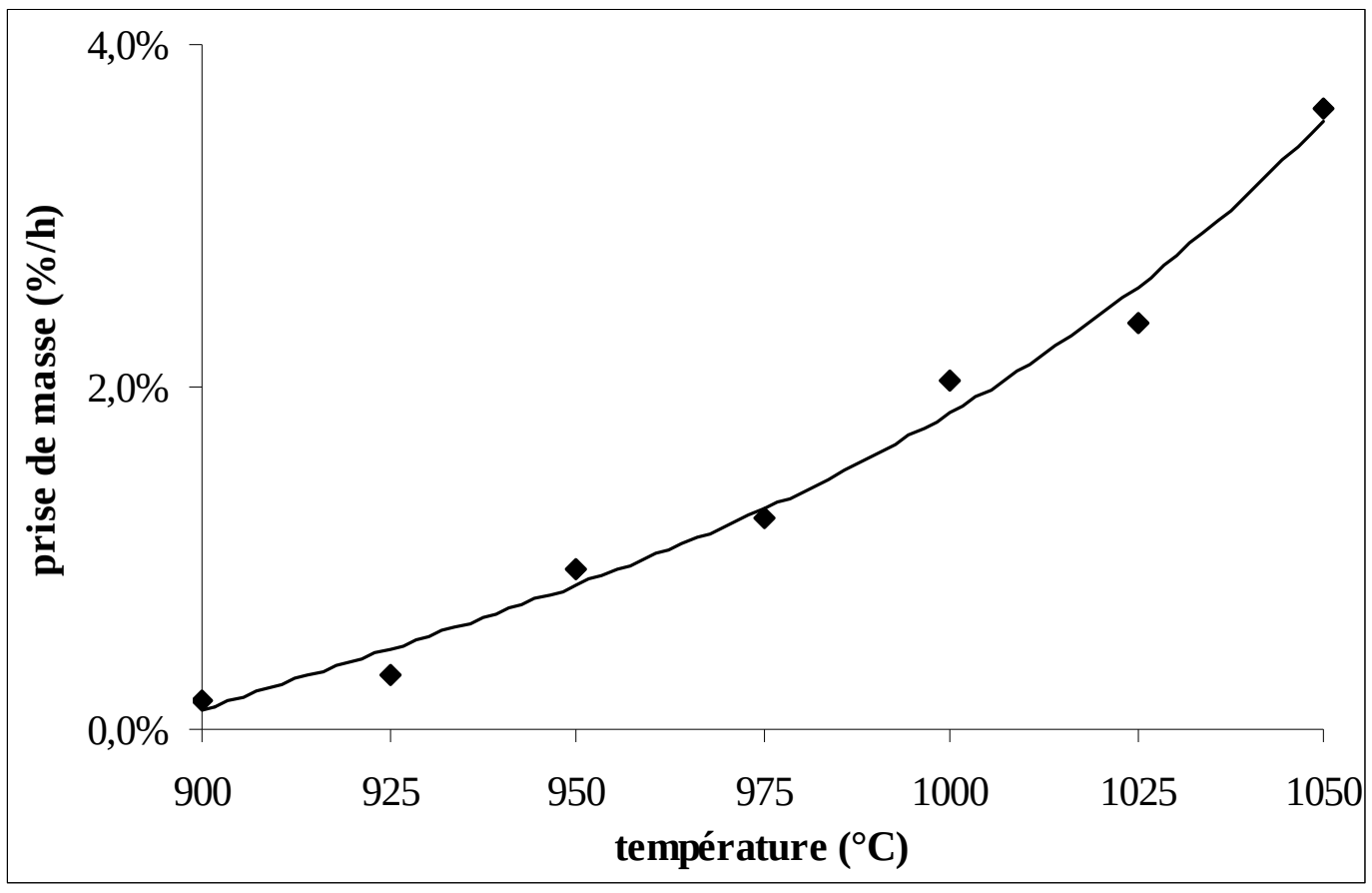

*Figure 3. Evolution de la prise de masse en fonction de la température*

## 3. Modèle de réaction

L'objectif est de développer un modèle détaillé de la réaction comportant un mécanisme homogène basé sur des processus élémentaires (Bounaceur et al., 2002) pour décrire les réactions en phase gazeuse de pyrolyse du propane, couplé à un mécanisme hétérogène décrivant les réactions de dépôt du pyrocarbone.

Ce modèle est construit en 3 parties qui ont été validées successivement :

- mécanisme de pyrolyse du propane conduisant à la formation des hydrocarbures aliphatiques et des aromatiques : benzène et toluène
- mécanisme des réactions des aromatiques et polyaromatiques (jusqu'au pyrène $C_{16}H_{10}$)
- mécanisme de formation du pyrocarbone.

### 3.1 Mécanisme de pyrolyse en phase gazeuse

Ce mécanisme décrit les réactions mises en jeu lors de la pyrolyse du propane et rend ainsi compte de la formation des différents hydrocarbures et molécules polyaromatiques observés. Il contient plus de 600 réactions essentiellement radicalaires faisant intervenir près de 180 espèces moléculaires et radicalaires. Des données cinétiques sont associées à chaque réaction, ce sont les paramètres de la loi d'Arrhénius-Kooij qui décrit l'évolution de la constante de vitesse d'une réaction en fonction de la température. Ces données sont majoritairement extraites de la littérature et peuvent tout aussi bien provenir d'études expérimentales faites dans des conditions opératoires différentes des nôtres (études d'oxydation, réacteur différents, réactifs différents…). Certaines constantes ont été calculées à partir du logiciel KINGAS, basé sur la théorie des collisions et sur des méthodes dérivant de la théorie du complexe activé (Bloch-Michel, 1995). Enfin pour les autres réactions, les paramètres ont été déterminés par des corrélations basées sur la relation entre la structure des espèces et leur réactivité, c'est à dire grâce aux analogies pouvant exister entre des réactions de même type mettant en jeu des molécules de structures proches. Dans la mesure où certaines réactions dépendent de la pression, leur constante cinétique est alors calculée à partir de la relation proposée par Troë, qui lie la constante de vitesse à une pression donnée, et les constantes extrapolées à pressions infinie et nulle (Troë, 1974). Enfin, les données thermodynamiques associées à chaque espèce du mécanisme (entropie, enthalpie de formation, capacité calorifique standard) sont essentiellement calculées par le logiciel THERGAS dont le principe repose sur les méthodes d'additivité de groupes ou dérivées de la mécanique statistique et proposées par Benson (Muller et al., 1995). Les autres données thermodynamiques non calculables à partir de ces méthodes proviennent de la littérature. Les simulations des réactions chimiques sont effectuées à l'aide du logiciel CHEMKIN II (Kee et al., 1993) dans des conditions identiques à celles des expérimentations. L'adéquation entre les résultats de modélisation et les résultats expérimentaux a permis de valider le mécanisme homogène de la pyrolyse du propane sur un large domaine de température (150°C) et de temps de passage (Ziegler et al., 2005b ; Ziegler et al., 2005c).

La validation du mécanisme de pyrolyse du propane a ensuite permis de réaliser une analyse de flux. Cette analyse permet de calculer les flux de production et de consommation de chacune des espèces du mécanisme et ainsi de construire les schémas réactionnels impliqués dans la pyrolyse du propane (Figure 4). Ces schémas permettent de mettre en évidence l'importance du rôle de certains radicaux dans la pyrolyse du propane. On citera en particulier l'importance des radicaux primaires issus du propane, $iC_3H_7$, $nC_3H_7$, $C_2H_5$. Ces trois radicaux gouvernent la répartition des voies de formation entre espèces en « $C_2$ » ou « $C_3$ ». En effet les radicaux $C_3H_7$ forment du propène qui à son tour sera à l'origine de la formation du propyne ($p\text{-}C_3H_4$) et du propadiène ($a\text{-}C_3H_4$). On notera le rôle important des premières réactions de décomposition du propane. D'autre part, le radical propargyle, $C_3H_3$, apparaît également comme un intermédiaire très important puisqu'il conduit à trois « familles » de produits ; à savoir les hydrocarbures à 4 atomes de carbone (« $C_4$ »), les dérivés du benzène et les dérivés du cyclopentadiène. On notera que le radical cyclopentadiényle issu du propargyle est lui-même une espèce importante car il est à l'origine de la formation du cyclopentadiène, de l'indène, puis du naphtalène, de l'acénaphtylène, du phénanthrène et de l'anthracène. Il intervient donc très fortement dans la formation des espèces polycycliques aromatiques organisées. A partir du schéma de flux, on remarque enfin que le benzène évolue plutôt vers des cycles aromatiques ramifiés tels le phényléthylène, l'éthylbenzène, le phénylacétylène ou le toluène.

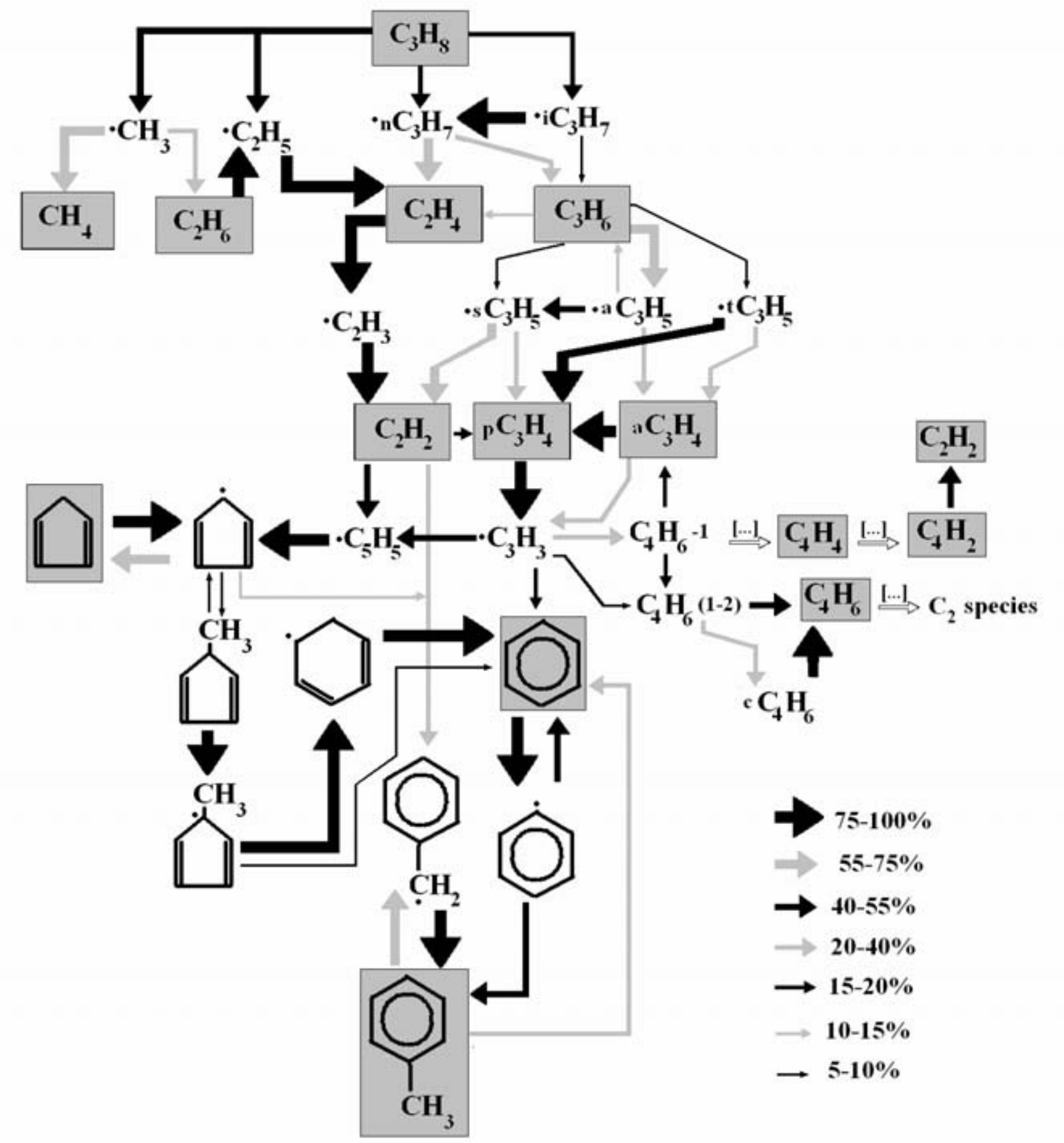

*Figure 4. Pyrolyse du propane: voies réactionnelles et importance relative des flux à 950°C et 1 s*

## 3.2 Mécanisme de formation du pyrocarbone

Une fois le mécanisme homogène validé, nous avons développé une approche globalisée, pour rendre compte du dépôt de pyrocarbone. Cette approche a consisté à représenter la réaction de formation du pyrocarbone comme une réaction globale en phase gazeuse de dépôt d'une espèce insaturée ou polyaromatique de la phase gazeuse (acétylène, éthylène, anthracène…). Les paramètres cinétiques ont été estimés à partir des résultats expérimentaux traduisant l'évolution de la masse déposée en fonction de la température, ainsi qu'à partir des différentes concentrations de la phase gazeuse. Cette réaction et ses paramètres cinétiques sont ensuite ajoutés au modèle homogène précédent qui est ensuite testé et validé par simulations. Il est alors possible de vérifier si une espèce peut expliquer ou non quantitativement le dépôt de pyrocarbone mesuré, tout en restant compatible avec les concentrations des autres produits de la phase gazeuse et ceci, en faisant abstraction du détail des processus chimiques mis en jeu lors de la formation de ce dépôt. Les conclusions obtenues par cette approche globalisée originale montrent que les petites espèces insaturées (telles que l'acétylène ou l'éthylène) peuvent rendre compte à la fois de la formation du pyrocarbone et de la composition du reste de la phase gazeuse. Par contre les réactions globalisées de dépôt des molécules aromatiques et polyaromatiques (par exemple : anthracène $C_{14}H_{10}$) conduisent à un dépôt très insuffisant et mènent, de plus à un très fort appauvrissement en ces espèces pourtant observées en sortie du réacteur (Ziegler-Devin et al., 2007). Nous avons donc retenu que les petites espèces insaturées (éthylène et acétylène principalement, cf. Table 1) seraient à l'origine du dépôt de pyrocarbone. Ces deux voies de formation de pyrocarbone ont donc été incorporées au mécanisme avec une pondération selon les constantes de vitesse d'addition sur un radical. Les résultats de simulation sont compatibles avec les dépôts mesurés (Figure 5) ainsi qu'avec les autres produits de la réaction (Figure 6).

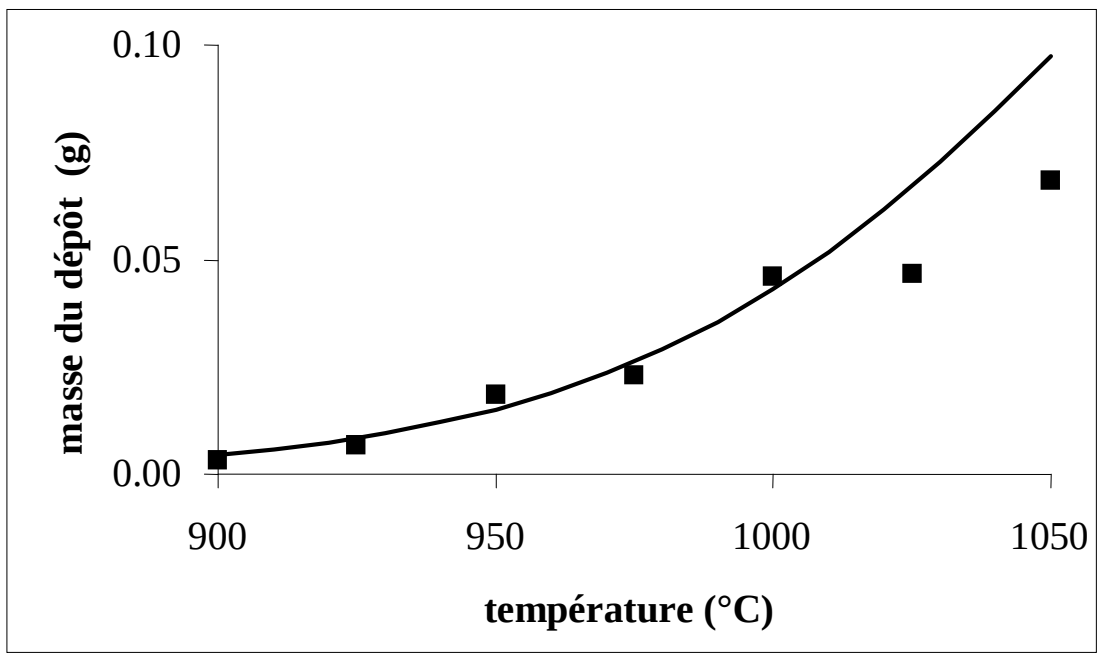

*Figure 5. Prise de masse en fonction de la température. Comparaison entre expériences (points) et simulation par dépôt d'éthylène et d'acétylène (ligne)*

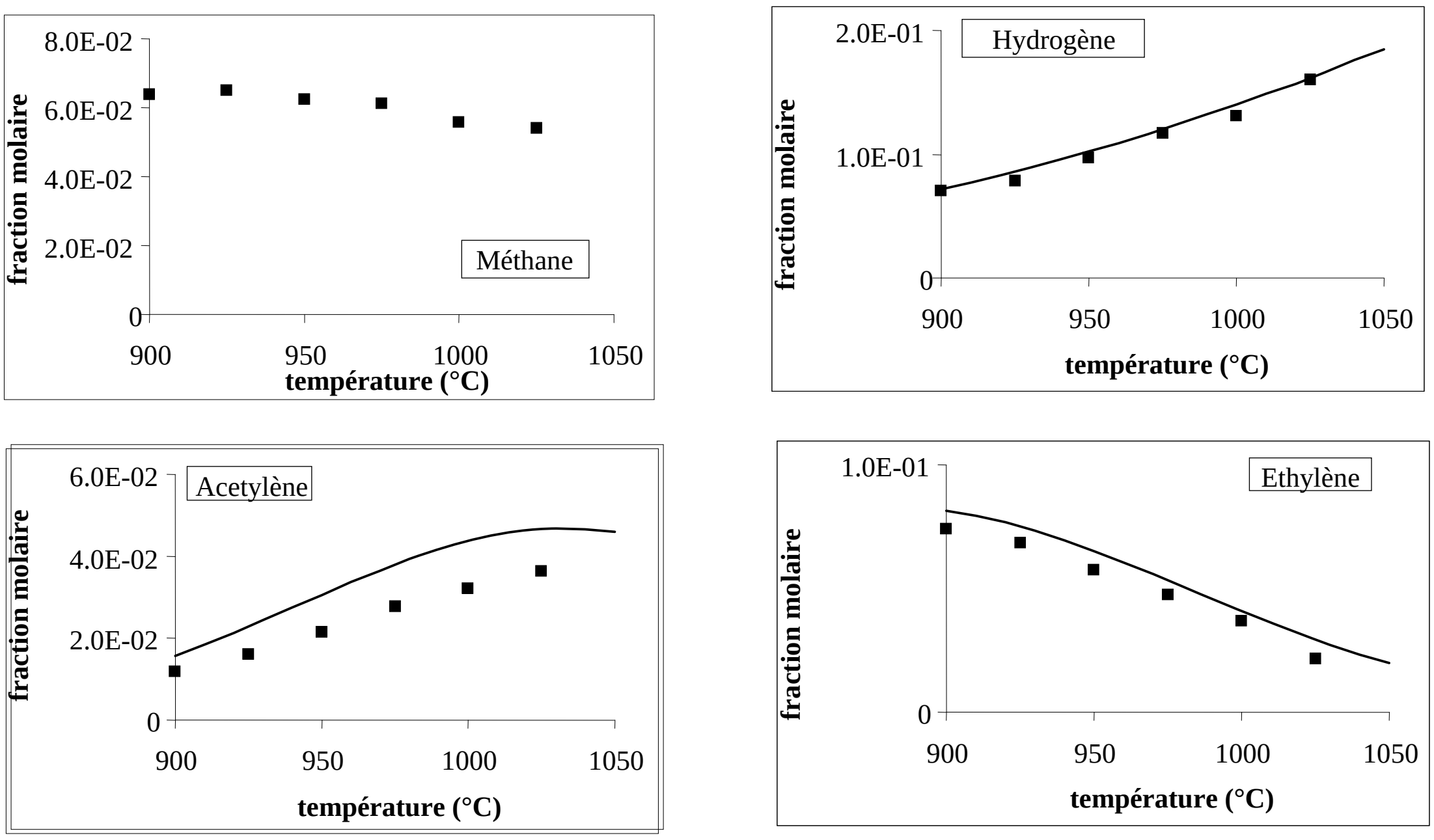

*Figure 6. Formation des produits légers en fonction de la température. Comparaison entre expériences (points) et simulation (ligne) par dépôt d'éthylène et d'acétylène*

## 4. Conclusions

La pyrolyse du propane à haute température (1000°C) et basse pression (<< 1 atm) conduit à la formation d'une faible proportion de pyrocarbone (< 10%) et de nombreux produits allant du méthane à de grosses molécules polyaromatiques. Une étude expérimentale détaillée, en dosant le maximum de produits de la réaction (30) dans un large domaine expérimental, a été réalisée en réacteur idéal parfaitement agité. Ces résultats nous ont permis de proposer et valider un modèle détaillé de la réaction comportant un mécanisme homogène basé sur des processus élémentaires pour décrire les réactions en phase gazeuse de pyrolyse du propane, couplé à un mécanisme hétérogène, décrivant les réactions de dépôt du pyrocarbone. Nous avons montré que les petites espèces insaturées (éthylène et acétylène principalement) seraient à l'origine du dépôt de pyrocarbone alors que les polyaromatiques tels que l'anthracène, ne peuvent rendre compte des résultats observés. Une perspective à ce travail consistera à développer un mécanisme détaillé hétérogène en s'inspirant des conclusions obtenues par cette approche globalisée, à savoir le dépôt préférentiel de petites espèces (acétylène, éthylène).

### Références

### Remerciements

Les auteurs remercient Snecma Propulsion Solide et en particulier, C. Descamps, C. Robin-Brosse, J.M. Rougès, J. Thébault.